\documentclass[pdflatex,sn-mathphys-num]{sn-jnl}

\usepackage{graphicx}%
\usepackage{multirow}%
\usepackage{amsmath,amssymb,amsfonts}%
\usepackage{amsthm}%
\usepackage{mathrsfs}%
\usepackage[title]{appendix}%
\usepackage{xcolor}%
\usepackage{textcomp}%
\usepackage{manyfoot}%
\usepackage{booktabs}%
\usepackage{algorithm}%
\usepackage{algorithmicx}%
\usepackage{algpseudocode}%
\usepackage{listings}%

\theoremstyle{thmstyleone}%
\theoremstyle{thmstyletwo}%

\theoremstyle{thmstylethree}%

\begin{document}

\title[Liquid-liquid phase separation at the
interface of an evaporating droplet; formation
of a regular lattice pattern]{Liquid-liquid phase separation at the
interface of an evaporating droplet; formation
of a regular lattice pattern}


\author*[1,2]{\fnm{Vahid} \sur{Nasirimarekani}}\email{vahid.nasirimarekani@ds.mpg.de}

\affil*[1]{\orgdiv{Max Planck Institute for Dynamics and Self-Organization,}
\normalsize{Am Fassberg 17, 37077, Göttingen, Germany}}

\affil[2]{\orgdiv{Laboratory of Fluid Physics and Biocomplexity},
\normalsize{Am Fassberg 17, 37077, Göttingen, Germany}}


\abstract{Evaporation alters the molecular interactions and leads to phase separation within the evaporating liquid. The question of whether evaporation could lead to specific phase separation at the liquid interface and eventually to the formation of patterns in small liquid volumes remains unaddressed. In this study, we investigated the liquid-liquid phase separation (LLPS) of an organic polymeric monomer in a salt-containing buffer within an evaporating sessile droplet. We observed that LLPS occurs at the dynamic interface of the droplet and leads to the formation of polymeric coacervates or regular lattice patterns depending on the initial concentration of the polymer. Our results show that the interaction of salt with the polymeric monomers at the droplet interface can lead to LLPS and the formation of regular patterns. This study suggests that the sessile droplet setup can be utilized to achieve very regular patterns as a result of LLPS.}

\keywords{liquid-liquid phase separation, evaporating droplet, microphase separation, pattern formation}



\maketitle
\newpage
\section*{Introduction}

\begin{figure}[h!]
    \centering
    \includegraphics[width=0.95\textwidth]{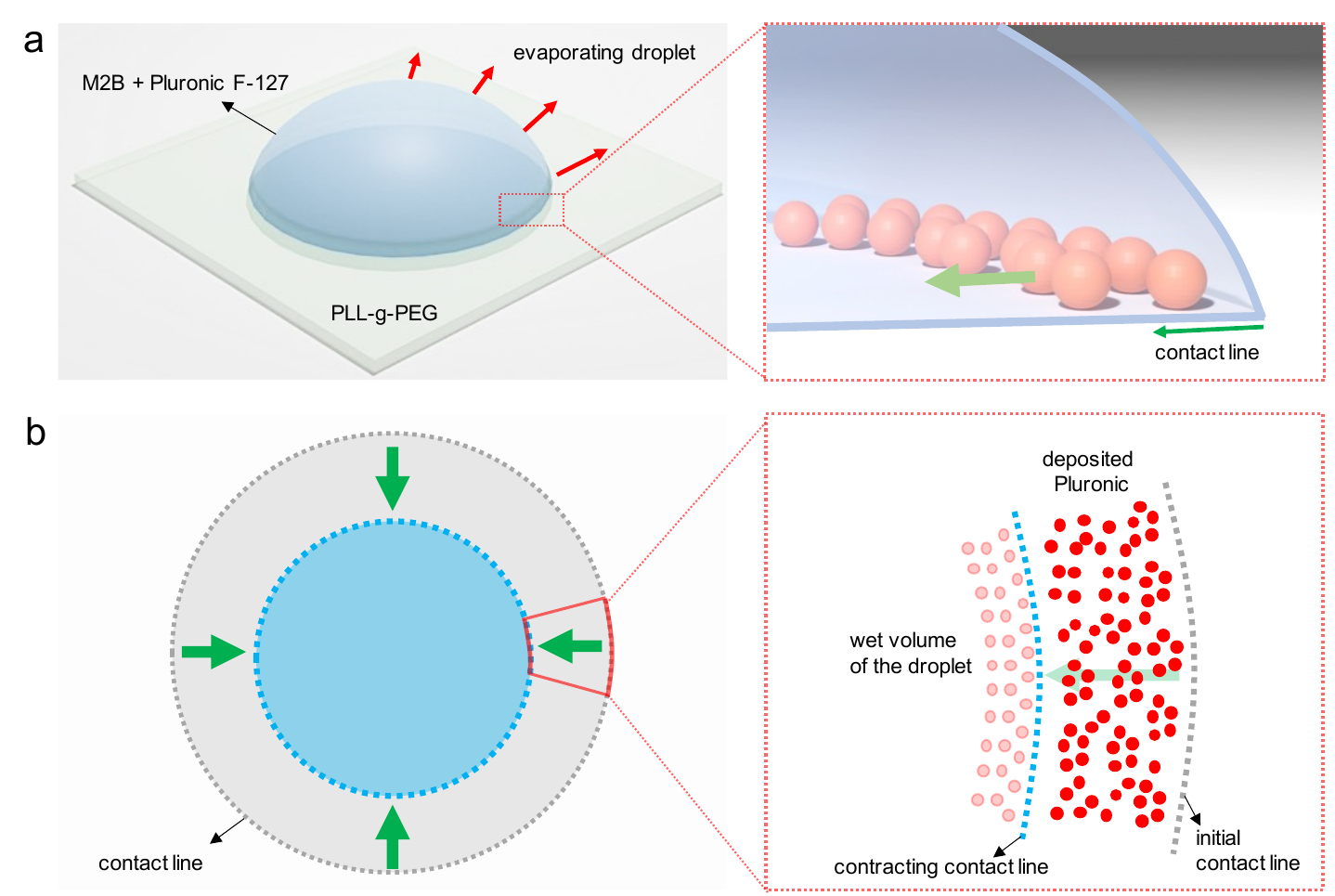}
    \caption{Schematic representation of the contracting sessile drop used to study the pattern formation of Pluronic polymers at the droplet interface. a) The overall shape of the droplet on a glass substrate, where the glass substrate is functionalized with a Pll-g-PEG surface to prevent pinning of the droplet and binding of the Pluronic molecules to the substrate. The inset view shows the presence of Pluronic molecules at the solid-water interface, which reduce the surface tension. b) A 2D drawing shows the contracting droplet moving towards the center of the droplet. The gray dash-line represents the initial contact line of the droplet, while the blue dash-line represents the contracting contact line. As the droplet contracts, Pluronic molecules will deposit on the substrate (shown as red dots in the inset view).}
    \label{fig:Figure1}
\end{figure}

The liquid phase may be the most complex phase of matter in the nature within which the life has formed and evolved. The dynamic molecular structure of fluids offers fluidity and facilitates the integration of other soluble and insoluble molecules. In other words, a fluid is a phase capable of accommodating various molecules, macromolecules, or solid particles, either in a random arrangement or in well-defined patterns. 
Soluble molecules dissolved in a liquid phase such as water may undergo liquid-liquid phase separation (LLPS), a phenomenon that occurs through coacervation, resulting in the formation of membraneless assemblies  \cite{hyman2014liquid,ghosh2021can}. LLPS occurs due to entropic or enthalpic effects within the liquid, ultimately leading to the formation of a new pattern of phase-separated molecules, specially in biological liquids~\cite{sing2020recent, quiroz2020liquid,mehta2022liquid}. 
\\
Reducing a liquid to smaller volumes can facilitate the LLPS by inducing the effect of physical parameters such as concentration effects and interfacial phenomena in a liquid. It also offers the possibility to observe the entire volume and how the liquid constituents undergo certain patterning to reach an equilibrium state. A droplet is a small volume of fluid that can form in a gas, liquid medium, or on a solid surface. A droplet sitting on a solid surface, a so-called sessile drop, enables the visualization of the entire volume of the liquid and its components~\cite{garcia2017sessile}. Therefore, it has been used as a setup to probe pattern formation with the possibility of applying a controlled environment around the droplet~\cite{weber1855mikroskopische,guo2021non,bell2022concentration}. 
\\
Evaporation is a naturally occurring physical process that applies to almost any volatile liquid exposed to the air. It induces various types of gradients within the liquid (concentration~\cite{bell2022concentration}, temperature~\cite{fang1999temperature}, surface tension~\cite{shereshefsky1936study} and etc.), which means occuring entropic and enthalpic changes in the liquid phase. Entorpic changes play a significant role in LLPS; therefore, evaporation of a volatile liquid would potentially result in phase separation of the soluble molecules~\cite{lohse2020physicochemical,othman2023liquid, nasirimarekani2023pattern}. 
A small drop of water containing salt and a surfactant would evaporate on a solid surface. Surfactants, molecules that reduce surface energy, are found in nature and specifically in seawater, where they play a crucial role in the formation of sea spray~\cite{frossard2019properties,gomez2011determination}. This raises the question of what kind of pattern a droplet containing surfactant and salt can leave on the surface during evaporation.
\\
It is known that the non-ionic surfactants tend to decrease the surface tension of an evaporating droplet~\cite{marin2016surfactant,shao2020role}; this means that the surfactant molecules flow to the interface and accumulate at the droplet-air interface (Figure \ref{fig:Figure1}.a). To understand the role of the interaction of salt ions with a nonionic surfactant in an evaporating liquid, we focus here on the pattern formation of a nonionic surfactant at the interface of an evaporating sessile droplet. More precisely, whether a dynamic droplet interface (shrinking droplet) can induce LLPS of the surfactant molecules at the contact line (Figure \ref{fig:Figure1}.b). 
We hypothesized that LLPS could occur at the interface of the droplet, potentially leading to the formation of regular patterns rather than a typical coffee ring~\cite{deegan1997capillary,yunker2011suppression} or isotropic deposition on the substrate. 
In this regard, we conducted experiments using a non-ionic surfactant, block copolymer, Pluronic F-127 (hereinafter referred to as Pluronic), in a salty buffer containing a mixture of mono- and multivalent salts adjusted to a biological pH level ($\sim 6.7$). 
Considering that concentration gradients near the droplet interface would play a role, as Pluronic tends to occupy the interface, we visualized the pattern formation at the interface of the droplet (Figure \ref{fig:Figure1}.b, inset view). 
\\
This article reports the experimental results of the described evaporating droplet to study LLPS at the droplet contact line and the resulting pattern deposition on the substrate. We report that we observe a regular lattice patterning of Pluronic on the substrate resulting from LLPS. Through a series of experiments and by reducing the complexity of the droplet, we show that the pattern is created by the interaction of salt and surfactant, which causes phase separation. We foresee 
the patterns of the non-ionic block copolymers could provide valuable insights into the pattern formation of biological macromolecules with similar block-like structures. This would also contribute in a broader context to the understanding of the role of evaporation in the origin of life on our planet~\cite{fox1976evolutionary,spruijt2023open}. 
\section*{Results and Discussion}
The evaporation of a droplet containing M2B (a salty buffer comprising 80 mM PIPES adjusted to pH = 6.9 with KOH, 1 mM EGTA, and 2 mM MgCl2), block copolymer (Pluronic F-127), and quantum dots (as tracer particles) was visualized using an inverted epi-fluorescent microscope (Figure S1). The quantum dots, which emit in the red wavelength range (585 nm) with an average diameter of 15-20 nm, were used to capture higher resolution images to visualize the resulting pattern.
\\
We investigated three different concentrations of Pluronic in the M2B buffer, namely, 0.5\%, 1\%, and 3\% w/w (weight basis). Considering that Pluronic has a critical micelle concentration (CMC) of around 0.7\%w/w~\cite{alexandridis1994micellization}, we selected concentrations both below and above the CMC to encompass a range of concentrations. Around 0.2 $\mu l$ of the mixture was pipetted on a functionalized glass surface, coated with Poly(l-lysine)-graft-poly(ethylene glycol) (PLL-g-PEG). The coating was chosen for its brush-like structure, which prevents the attachment of particles and organic or inorganic polymers to the glass surface~\cite{feng2018polymer, benetti2019using}. The formed droplet was dried out under the room temperature (20$^{\circ}$C, Relative humidity (RH) 35\%), and dried over time and the droplet contact line (the periphery of the droplet) was visualized and time-lapse images were recorded in order to capture the deposited pattern of the polymer on the glass substrate (Figure S1). 
\begin{figure}[h!]
    \centering
    \includegraphics[width=\textwidth]{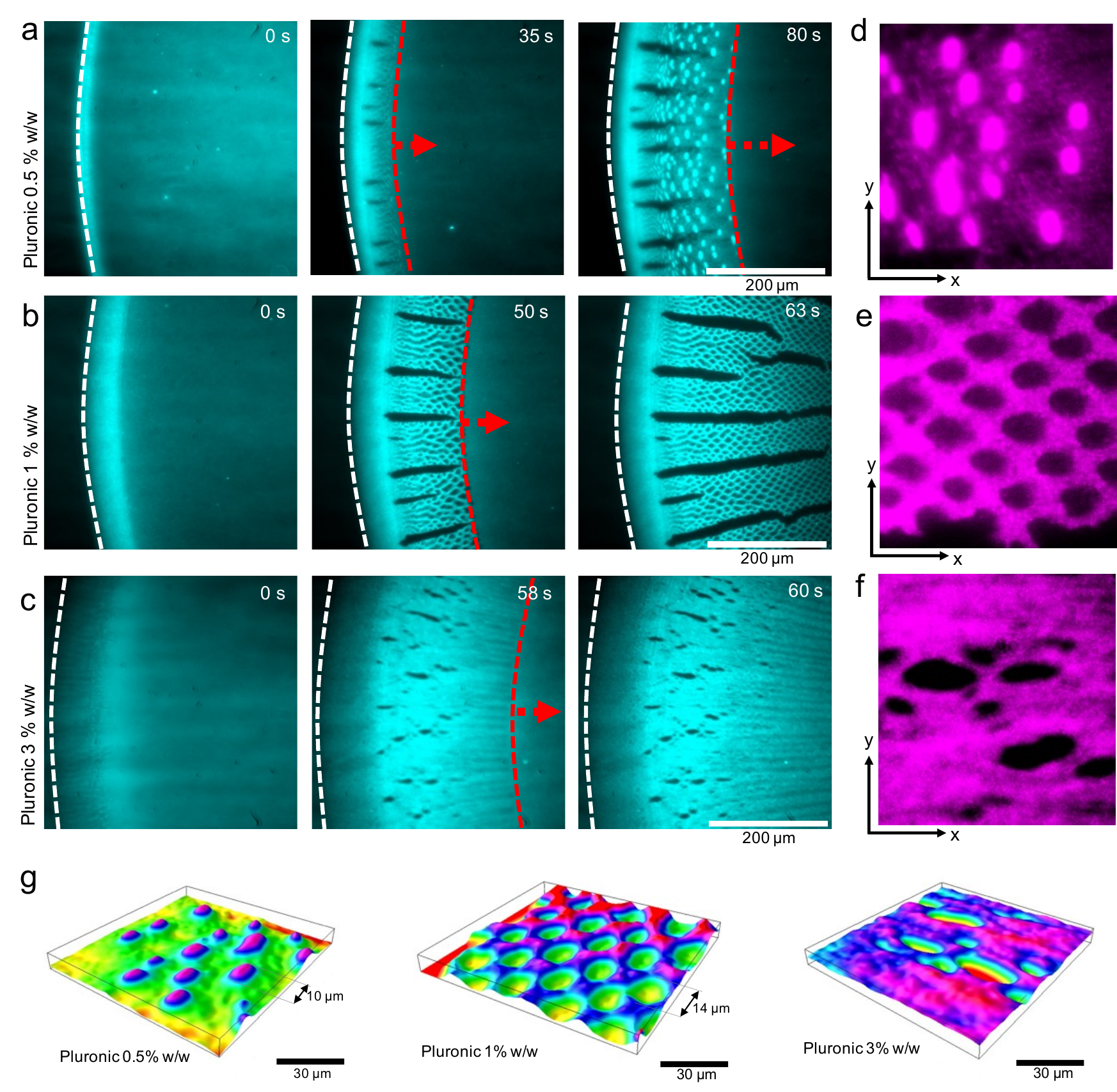}
    \caption{Effect of Pluronic concentration on pattern formation at the solid-water-air interface of the evaporating droplet. a-c) Time-lapse images of the droplet containing 0.5\%, 1\% and 3\% w/w, respectively. The increment of the Pluronic concentration shows transition of the pattern from cluster (0.5\%) to a lattice pattern (1\%) and to a continuous sheets with radial finger-like defects (3\%). d-g) views of the individual droplets containing the above concentration provide an enlarged view of the sample. g) 3D surface topology plot of the inset views in sections d, e and f.}
    \label{fig:Figure2}
\end{figure}
\subsection*{Initial concentration defines the final pattern formation at the contact line of an evaporating droplet}
The microscopy images depict that the initial concentration of the polymer determines the final patterning inside the evaporating droplet. As a function of the Pluronic concentration, the results show three distinct pattern formation inside the evaporating droplet: (i) discrete aggregates forming clusters with a random order, (ii) a regular lattice pattern, and (iii) a continuous sheet with radially aligned defects, respectively for Pluronic concentrations of 0.5\%, 1\% and 3\% w/w (Figure~\ref{fig:Figure2}).
\\
A closer look at the pattern of the 0.5\% droplet shows the deposition of dense clusters with a thin layer around the clusters (Figure \ref{fig:Figure2}. a,d). The clusters formed at the interface (in contact with air), with an irregular distribution along the contact line (Figure \ref{fig:Figure2}. a, 180s, the red dash line represents the droplet contact line). The density of the clusters decreases along the radial direction towards the center of the droplet. In addition, the radial rips (long radial defects) are seen perpendicular to the droplet contact line. 
\\
In the case of 1\% droplet, a single sheet of polymeric network can be seen with a clear lattice pattern and longer and clearly visible radial rips (Figure \ref{fig:Figure2}.b,e). The lattice pattern shows regular and linear alignments that cover a larger surface area of the droplet. By further increasing the Pluronic concentration to 3\% w/w, a continuous sheet with less visible rips, which can be seen more as finger-like radial defects, are observed (Figure \ref{fig:Figure2}.c,f). The 3D surface topology of the patterns was plotted based on the light intensity values, underlining the above observations (Figure \ref{fig:Figure2}.g). 
\\
The experiments in Figure~\ref{fig:Figure2} show that the initial concentration of the polymer significantly influences the final pattern of the evaporating droplet. The remarkable aspect is the formation of regular lattice patterns, which indicates that an adjustment of the initial concentration can lead to the formation of very regular patterns and 3D structures. Although it has already been reported that the initial polymer solution affects the size of the final features~\cite{bormashenko2006self,tsige2005solvent,schaefer2015structuring}, here we observe a transition from randomly arranged polymer coacervates to a regular lattice pattern indicates as a function of the initial concentration.
\\
It is noteworthy that the pattern formation occurred at the interface of the droplet with the solid substrate and the air (Figure~\ref{fig:Figure1}). As the droplet dried out, the contact line of the droplet moved towards the center, leaving behind a deposition of the polymer material (Video S1, S2). Microscopy was performed with quantum dots mixed with the polymer solution to obtain visible images of the final pattern. To understand the effects of the added quantum dot particles on the final pattern formation, we investigated the pattern formation of the polymer solution in the absence of quantum dots. When comparing two cases, no recognizable changes in the final pattern were observed (Figure S2). It was therefore assumed that the quantum dots do not influence the pattern formation of Pluronic. 
\subsection*{Liquid-liquid phase separation at the interface of the evaporating droplet}
\begin{figure}[h!]
    \centering
    \includegraphics[width=0.88\textwidth]{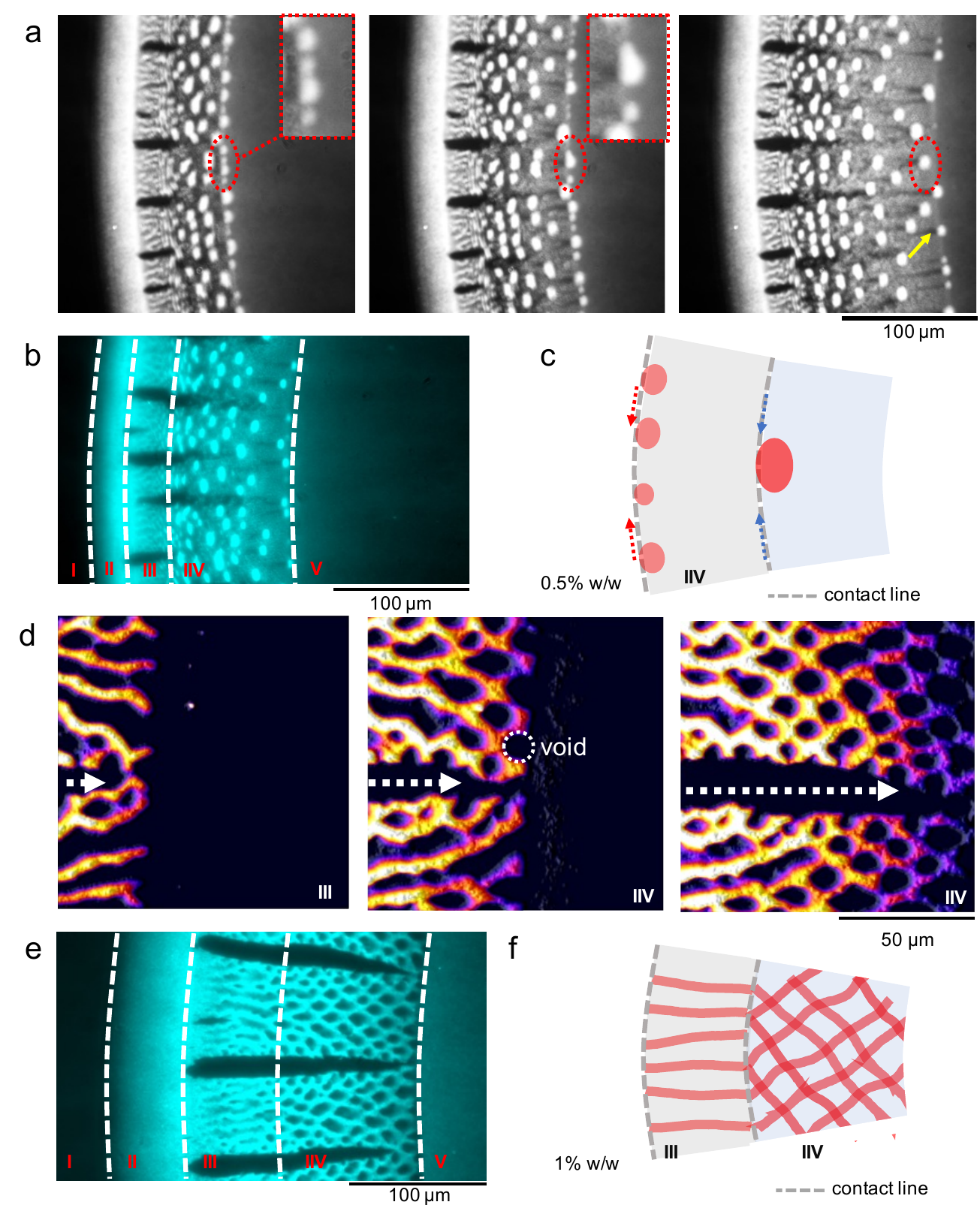}
    \caption{A closer view of the pattern formation of Pluronic at the dynamic interface of the evaporating droplet. a) Experimental observation of the LLPS and the formation of polymer clusters at the interface of the droplet. The smaller clusters merge and combine to form larger clusters due to the contraction and dynamic nature of the contact line. b) The pattern formation for 0.5\% Pluronic droplet in different regions from the contact line. c) Schematic representation of the merging of polymer coavates into larger clusters. d) Experimental results of the microphase separation of Pluronic at the contact line of the evaporating droplet, which leads to the formation of an initially finger-like pattern and then a regular lattice pattern. The occurrence of voids at the interface of the drop in the case of a lattice pattern is emphasized. e) The pattern formation for 1\% Pluronic droplet in different regions from the contact line. f) The schematic representation show that the lattice pattern is formed as a transition of finger-like structures. Different regions are distinguished at the contact line of the drop, such as: (I) glass coating, (II) coffee ring area, (III) finger-like pattern, (IIV) clusters or lattice pattern and (V) the central part of the droplet.}
    \label{fig:Figure4}
\end{figure}

The microscopic pattern formation in the 0.5\% Pluronic droplet is the result of LLPS at the interface of the droplet. This LLPS is a function of the concentration of the polymer material, the salt and the pH value~\cite{imae1988liquid,narayanan1999salt}. Evaporation reduces the droplet volume which results in increment of polymer concentration. Therefore, the phase separation of polymeric monomers is seen as a coacervate clusters at the contact line (the monomers occupy the interface due to their surfactant behavior~\cite{almohammadi2023evaporation}). The clusters then pushed together and form bigger ones and the rest of the monomers which do not coacervate are deposited as a thin layer (Figure~\ref{fig:Figure2}.a,d). 
Time-lapse images show that polymer clusters form as smaller aggregates at the contact line, which transform into larger clusters by merging several smaller clusters as the contact line proceeds towards the droplet center. An example case in which 4 smaller aggregates cluster to a bigger one as the the interface pushes them together along the interface (highlighted in a Figure~\ref{fig:Figure4}.a,c). In addition, the larger clusters are carried further by the contact line, which can be recognized by the swiped tail footprint of the clusters (Figure \ref{fig:Figure4}.a, yellow arrow points the swiped footprint). 
\\
The lattice pattern is a specific form of LLPS that occurs by a higher concentration of the polymer(Figure~\ref{fig:Figure4}.d). So-called micophase separation is an spontaneous form of phase-separated structures with microscopic length scales due to intramolecular phase separation in concentrated solutions~\cite{lohse1997microphase, lee2023polymerization,chakrabarti1989microphase}. Microphase separation in Pluronic polymers refers to the spontaneous organization of their molecular components into distinct microscopic domains within the polymer matrix. This phenomenon arises from the amphiphilic nature of Pluronic molecules, which possess both hydrophilic and hydrophobic segments. The segregation of the phase separated segments leads to the formation of well-defined regions or phases within the polymer~\cite{teran2014thermodynamics,thelen2014phase}.  
Moreover, this self-assembly behavior is driven by thermodynamic considerations such as the minimization of free energy at the interface~\cite{likhtman1997theory}. 
Hence, the parallel finger-like radial phase separation is seen in region III (in both concentrations, Figure~\ref{fig:Figure4}.b,e) can be explained as a swelled microphase separation of Pluronic over the CMC. We suggest that contraction of the interface as a physical factor and increasing the w/w concentration of the Pluronic throughout the course of evaporation results in transition of the microphase separation from linear finger-like structures to the lattice pattern (Figure~\ref{fig:Figure4}.e,f, transition from region III to IIV).
\\
The lattice pattern formation in 1\% droplet is dominantly influenced by appearing of the voids at the interface and the radial rips  (Figure \ref{fig:Figure4}.d,e, white arrow represents the rips). The periodic pattern of the voids along the interface results in interconnecting lattice pattern. During formation at the interface, the voids have a circular geometry (Figure \ref{fig:Figure4}.d).
On the other hand, the rips form in the zone II and simultaneously grow over time along the droplet as far as the lattice pattern is formed (Figure \ref{fig:Figure4}.d). The rips were also visible in the 0.5\% case, but they are more dominant in 1\% case. The rips seem to cut through the lattice sheet forming radial cracks which do not disturb the lattice pattern (Video S3). 
\\
The lattice pattern shows a thin layer sheet-like material property. Initial concentration of 1\% droplet is above the CMC of Pluronic. Hence, the droplet contains stable micelles and free monomers simultaneously. The micelles have a hydrodynamic radius 1 order of magnitude larger than the radius of gyration of the unimers~\cite{xiang2016concentration,bate2019collective}. This gives a gel-like material to Pluronic which was also shown in our previous work, mixed with biological bio-polymers~\cite{nasirimarekani2021tuning}. Therefor, the continuous sheet and formation of rips is related to gel-like material property of the Pluronic mixture. In addition, the rip defects in a complimentary experiment was seen to push apart micometer sized solid particles embedded inside the polymeric material (Figure S3).
\\
Here, we showed that the reported final patterns of the Pluronic droplet form due to LLPS at the interface of the droplet, which is directly depends the initial concentration on the polymer inside the evaporating droplet. We believe that in addition to the influence of polymer concentration, salt and ionic strength is a dominant factor driving the system to phase separation. Therefore, the question of the role of salt in the observed LLPS and its final patterning is raised in the following section.  
\subsection*{LLPS is obtained by the salting-out effect at the interface of the evaporating droplet}
\begin{figure}[h!]
    \centering
    \includegraphics[width=\textwidth]{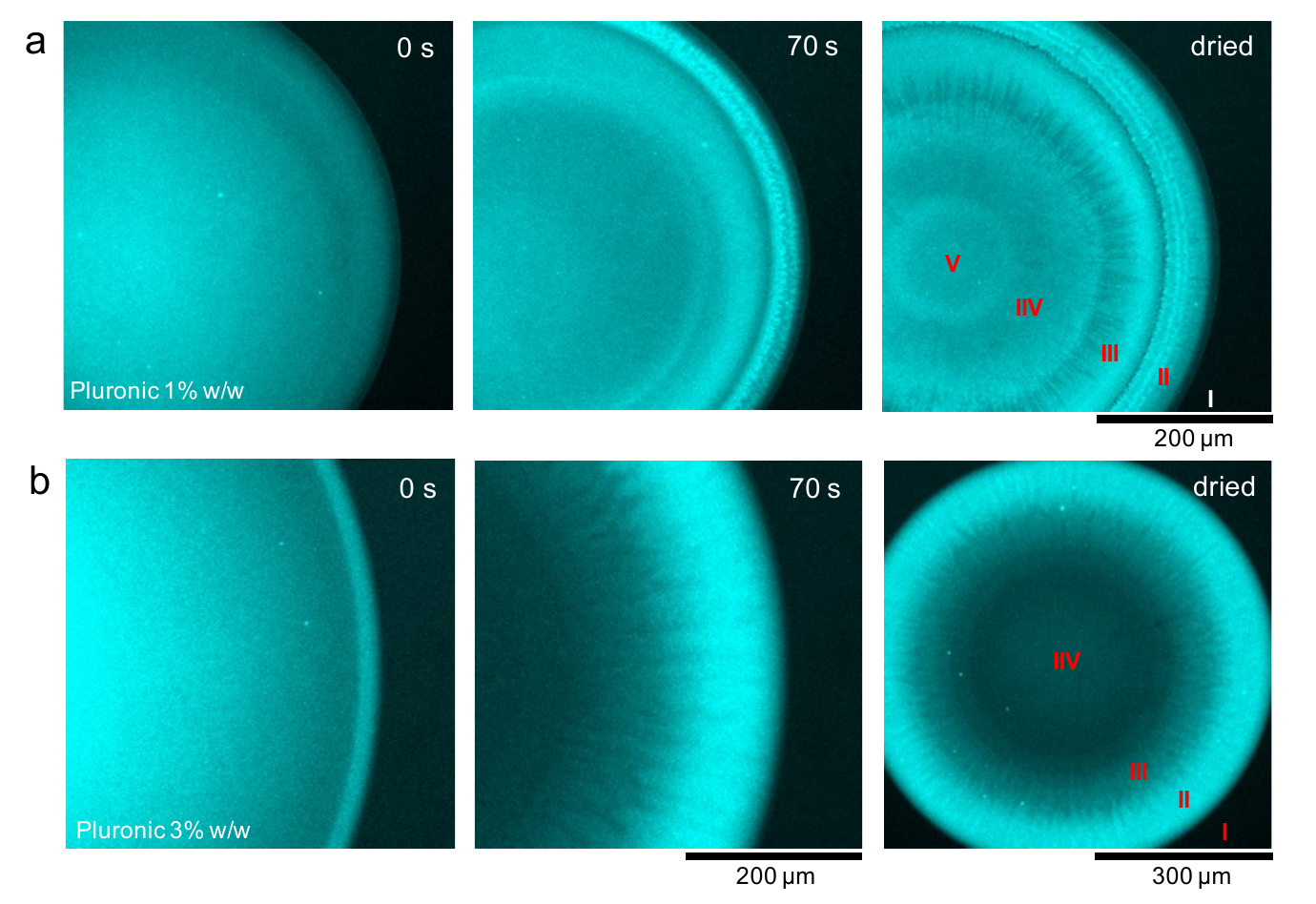}
    \caption{Pattern formation of Pluronic block co-polymer in water (absence of salt) inside an evaporating sessile droplet. a,b) Time-lapse images of droplet containing Pluronic concentration of 1\% and 3\% w/w, respectively. No LLPS separation and regular pattern formation or radial defects is observed.}
    \label{fig:Figure5}
\end{figure}

To experimentally observe the role of salt in LLPS on the reported patterns, further experiments were performed without salt (in mQ water as buffer) using the same setup as in the previous sections. In this regard, two different concentration of Pluronic as 1\% and 3\% w/w were tested.
\\
The results show that in the absence of salt, no LLPS, coavervate or lattice pattern formed. In contrast, a very distinct deposition compare to previous results is observed (Figure~\ref{fig:Figure5}). The main differences are the absence of any regular microscopic pattern and presence of the rips. Although some radial alignments can be observed, no clear regular pattern can be recognized (Figure~\ref{fig:Figure5}.a, dried). In 1\% droplet, 5 different regions, analogous to the previous results in Figure~\ref{fig:Figure4}.e were observed. Increasing of Pluronic concentration to 3\% w/w resulted in more continuous sheet with no visible radial defects (Figure~\ref{fig:Figure5}.b). 
\\
Salt tends to reduce the unspecific polymer interaction with water molecules through hydrogen bonding, which is known as salting-out effect~\cite{endo2012salting,grover2005critical,hey2005salting}. Hence, it increases the monomeric interaction in the polymer phase which promote the phase separation~\cite{imae1988liquid}. Thermodynamically speaking, presence of salt causes depletion effect as the result of entropic forces~\cite{aseyev2011non} due to the repulsive ionic interaction in between salt ions and Pluronic monomers (non-ionic block co-polymer). This results in a different phase behavior of Pluronic polymers, including their propensity for phase separation. Therefore, salt interaction with the polymer is the main mechanism for LLPS at the interface of the evaporating droplet. 
\\
Salt also impacts the rheological properties of Pluronic solutions, such as viscosity and elasticity~\cite{russo2024tuning}. Changes in salt concentration influences the interactions between polymer chains, affecting the overall flow behavior and mechanical properties of the solution over the course of drying~\cite{pandit2000effect}. In this regard, presence of the rips in previous experiments highlights the change of mechanical properties of the final deposited Pluronic layer. 
\\
Changes in salt concentration or composition can affect the balance of interactions between the polymer segments, leading to changes in the size, shape and stability of the microphase-separated domains~\cite{liu2000salt}. It is worth noting that the salt composition and concentration in this work was chosen to have a biological pH; therefore, a mixture of polyvalent and monovalent salts was used. We hypothesize that the observed LLPS and pattern formation can be manipulated or tuned by the type and concentration of salt. Such a study is beyond the scope of this work, but would be essential if some degree of control over the size and periodicity of the lattice pattern is desired. 
\subsection*{The phase separation influences the droplet contraction and the evaporation rate}
\begin{figure}[h!]
    \centering
    \includegraphics[width=\textwidth]{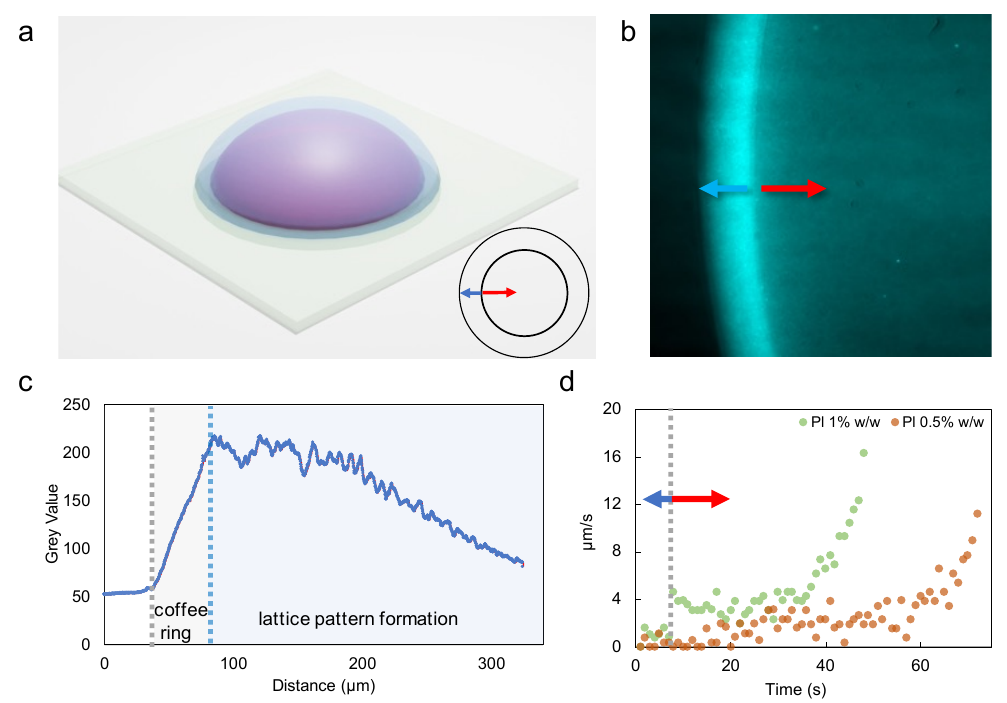}
    \caption{The dynamics of the droplet interface. a,b) Schematic representation and microscopy image highlighting two different dynamics observed at the droplet interface. The blue arrow stands for the coffee ring region and the red arrow for the inward contraction dynamics in which the pattern formation takes place. c) The gray value measurements of the drop represent the light intensity of the polymer solution, in the coffee ring and the lattice pattern formation regions. d) Plot of the velocity measurements of the droplet contact line for 0.5\% and 1\% w/w Pluronic droplets.}
    \label{fig:Figure3}
\end{figure}

Two different dynamics are observed at the droplet contact line of the evaporating droplet. First, an outward flow toward the droplet contact line which carries the polymeric material to the contact line in the early stages of the drying process, (Figure~\ref{fig:Figure3}.a,b blue arrow). The outward flow results in formation a ring shape similar to typical coffee ring~\cite{deegan1997capillary,yunker2011suppression}. Second, upon reaching a transition point, an inward contraction is observed towards the center of the droplet (Figure~\ref{fig:Figure3}.a,b red arrow). This reversal is particularly visible in 0.5\% and 1\% w/w Pluronic concentrations. In the case of the lattice pattern, the gray value measurements (light intensity) show that the transition to inward contraction occurs at the peak intensity, which is detected by the deposited polymers (Figure \ref{fig:Figure3}.c). In addition, the slope of the light intensity over time shows that the polymer deposition in the coffee ring region increases linearly. This suggests that LLPS occurs when the concentration of the polymer reaches a certain threshold value that favors phase separation. 
\\
The inward contraction of the contact line towards droplet center occurs due to the phase separation of polymeric monomers and the pattern formation. Tracking the contact line over time reveals that LLPS alters the contraction dynamics of the droplet in big extend. The velocity measurement of the contact line (with reference to the initial contact line of the droplet) shows that the contact line in the inward contraction region moves faster compared to the coffee ring region (Figure \ref{fig:Figure3}.d). There is a clear jump in the velocity values once the grid pattern begins to form. Since the contraction dynamics of a sessile droplet are directly related to the evaporation rate, pattern formation leads to a higher evaporation rate due to phase separation of the Pluronic molecules and facilitates evaporation~\cite{othman2023liquid, kalume2018liquid}. Comparing two different concentrations of 0.5\% and 1\% shows that for a similar droplet size, the droplet with the lattice pattern (1\% droplet case) dries out 1.6 times faster than 0.5\% droplet case. In addition, the velocity values in the inward contraction region tend to increase as the contact line of the droplet moves towards the center of the droplet. This is due to the reduction in droplet volume, which increases the contraction ratio due to a larger surface-to-volume ratio~\cite{saenz2015evaporation, cazabat2010evaporation}. 
\\
It is important to note that in droplets contaning surfactant, the Marangoni flow also leads to an inward flow resulting from the surface tension gradient at the contact line, which tends to lead to an inward flow towards the droplet center~\cite{hu2006marangoni,hu2005analysis}. In other non-ionic surfactants such as polyethylene glycol (PEG), Marangoni flow results in strong contraction of the droplet~\cite{kim2022multiple,seo2017altering}. Nevertheless, Pluronic concentrations which are used here result in a dense polymeric mixture and the pattern formation dominates the contraction of the droplet. 

\section*{Conclusion}
In this work, a series of experimental results on the observation of LLPS at the interface of an evaporating droplet were presented. We have shown that evaporation can lead to the formation of a regular lattice pattern of Pluronic polymer. The initial concentration of the polymer was found to be the dominant factor for the final pattern formation. In addition, ionic repulsion and salting-out effects were the main mechanism for triggering phase separation. The observation of a regular lattice pattern suggests that the evaporating droplet has the potential to generate regular microstructural patterns. However, to adjust the size and 3D structure of the lattice pattern, the effects of salt concentration and evaporation parameters such as temperature and relative humidity need to be investigated. 
\newpage
\section*{Materials and Methods}
\subsection*{Material}
\textit{Pluronic:} Pluronic® F-127 in powder form and suitable for cell culture was purchased from Sigma Aldrich, Germany. The powder was dissolved in M2B buffer or water and autoclaved prior to be used. Pluronic is an amphiphilic triblock copolymer with a central hydrophobic chain of polyoxypropylene (poly(propylene oxide)) and two hydrophilic chains of PEG.
\\
\textit{Quantum dots:} Qdot™ 585 ITK™, carboxyl quantum dots were purchased from ThermoFishers, Germany. The quantum dots were mixed in the M2B buffer or MQ water in 20 pM concentration. 
\\
\textit{Glass slides:} High-resolution glass slides were used to achieve better optical properties. The glass slides were purchased from Paul Marienfeld GmbH \& Co. KG, Germany in standard size, 24*60 mm. 
\\
\textit{Beads:} The 1 $\mu m$ sized carboxylate-modified polystyrene latex micro-spheres were purchased from Bangs Laboratories, Inc.
\subsection*{PLL-g-PEG coated glass surface preparation}
Microscope glass slides were cleaned by washing with 100\% ethanol for 10 mins and rinsing in deionized water. They were further sonicated in acetone for 30 min and incubated in ethanol for 10 min at room temperature. This was followed by incubation in a 2\% Hellmanex III solution (Hellma Analytics) for 2 h, extensive washing in deionized water, and drying with a filtered dry airflow. The cleaned coverslips were then surface activated in
oxygen plasma (FEMTO, Diener Electronics, Germany) for 30 s at 0.5 mbar. 0.1 mg/ml Poly(L-lysine)-graft-poly(ethylene glycol) (PLL-g-PEG) (SuSoS AG, Switzerland) in 10 mM HEPES (pH 7.4, at room temperature) was poured onto the glass cover slips and incubated for 30 mins. The coating solution was then dried out with a filtered dry airflow.
\subsection*{Preparation of the droplet solution}
M2B buffer (80 mM PIPES, adjusted to pH = 6.9 with KOH, 1 mM EGTA, 2 mM MgCl$_{2}$) was used as the basic buffer to dissolve Pluronic. 0.5\%, 1\% and 3\% w/w (weight basis) of Pluronic solutions were prepared by simply adding the Pluronic powder inside the M2B buffer and mixing on a magnetic stirrer. For 100 $\mu l$ volume of Pluronic solution 1 $\mu l$ of quantum dots were added as tracers particles. Moreover, in the case of micron-sized particles embedded in lattice pattern, 1 $\mu m$ beads were added in the corresponding Pluronic solution (in M2B).  For the mQ water experiments Pluronic was dissolved in mQ water as was explained for M2B buffer.
\subsection*{Evaporating droplet experiments}
A 0.2 $\mu l$ of the solution was pipetted on the PLL-g-PEG coated glass slide, which was then positioned on the microscope stage. A transparent box with the size of 5*5*2 cm was positioned on top of the glass slides to avoid any strong air flow over the droplet which could result in heterogeneous drying of the droplet.  
\subsection*{Image Acquisition and analysis}
\textit{Image acquisition:} Image acquisition was performed using an inverted fluorescence microscope Olympus IX-71 with various objectives from 4$\times$ objective (Olympus, Japan) to 60$\times$ oil-immersion (Olympus, Japan), depending on the experimental setup. For excitation, a Lumen 200 metal arc lamp (Prior Scientific Instruments, U.S.A.) was applied. The images were recorded with a CCD camera (CoolSnap HQ2, Photometrics). The frames were acquired with a variable rate according to the experiment with an exposure time of 200 ms for a variable time according to the experiment.
\\
\textit{Image analysis:} ImageJ software was used for the analysis of the acquired images. Manual tracking plugin was used to track the interface velocity and the data were plotted in Microsoft Excel. 
\section*{Acknowledgments}
The author acknowledges the critical and valuable comments of Prof. Bodenschatz. Additionally, I would like to thank Katharina Gunkel and Lasse Lehmann for their assistance in preparing the material. This project was supported by funding of Max Planck Society and the Klaus Tschira Boost Fund, a joint initiative of the German Scholars Organization and the Klaus Tschira Foundation.
\section*{Author Contributions}
As a single author, the entire work, from the conception and execution of the experiments to writing the manuscript, was carried out by the author himself. 
\section*{Competing Interests}
The authors declare no competing interests.
\bibliography{sn-bibliography}

\newpage
\section*{Supporting Information}
\setcounter{figure}{0} 
\renewcommand{\figurename}{Figure S.}
\subsubsection*{Experimental setup}
\begin{figure}[h!]
    \centering
    \includegraphics[width=0.8\textwidth]{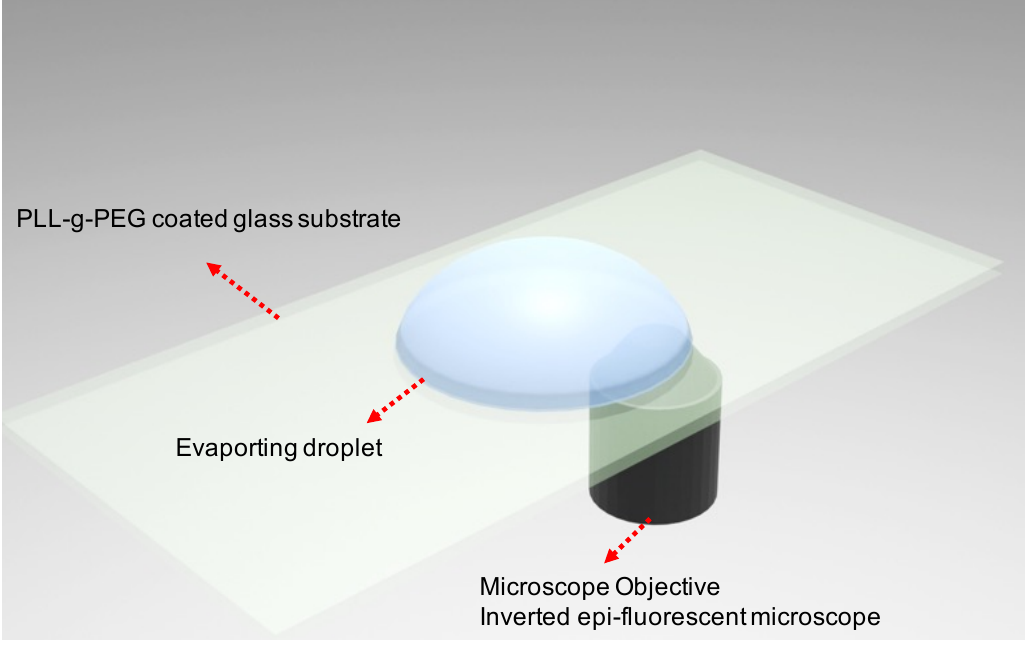}
    \caption{3D schematic of the experimental setup for investigating pattern formation at the interface of an evaporating droplet.}
    \label{fig:FigS1}
\end{figure}
\newpage
\subsubsection*{Lattice pattern formation in polarized light}
The formation of the lattice pattern in the absence of quantum dos was imaged using polarized light microscopy to determine the influence of the quantum dots on the final pattern formation. 
\begin{figure}[h!]
    \centering
    \includegraphics[width=0.8\textwidth]{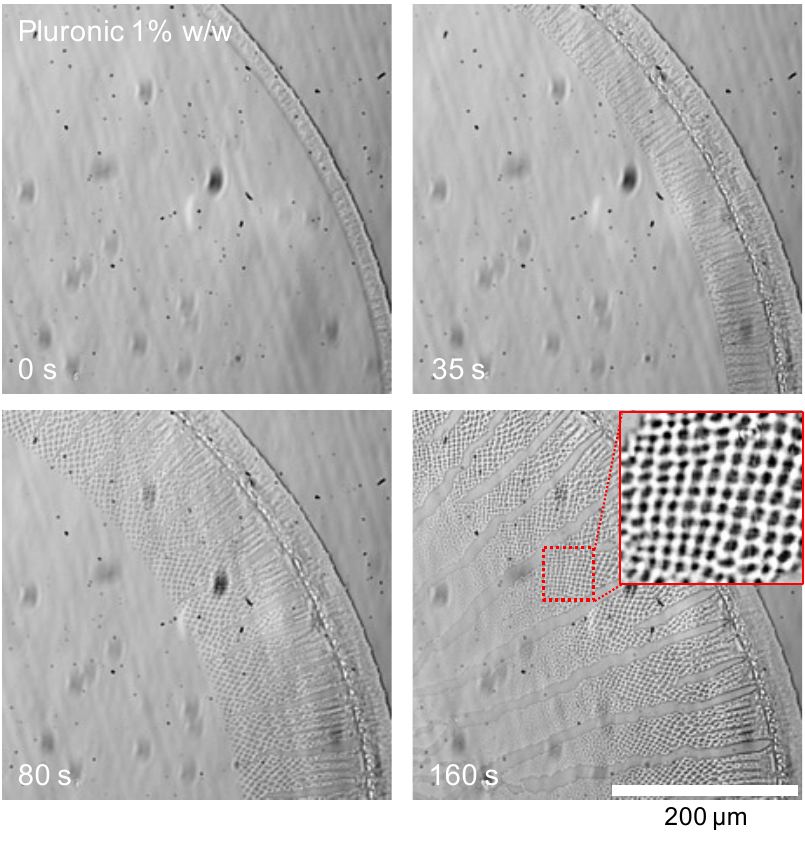}
    \caption{Polarized light microscopy of the lattice pattern formation without quantum dots. The inset view highlights the formation of the lattice pattern and shows that the quantum dots do not interfere with the formation of the microscopic pattern inside the Pluronic droplet.}
    \label{fig:FigS1}
\end{figure}
\newpage
\subsubsection*{Gel-like mechanical properties of the Pluronic film}
We questioned whether larger particles would remain in the Pluronic network because the quantum dots are nearly weightless. 1 $\mu m$ sized particles were mixed with Pluronic+M2B buffer (1\% w/w concentration of Pluronic) and polarized and fluorescence microscopy was performed to observe the pattern and particles within the evaporating droplet. The results lateral movement of the beads where the rips and the lattice pattern are formed. 
\begin{figure}[h!]
    \centering
    \includegraphics[width=0.8\textwidth]{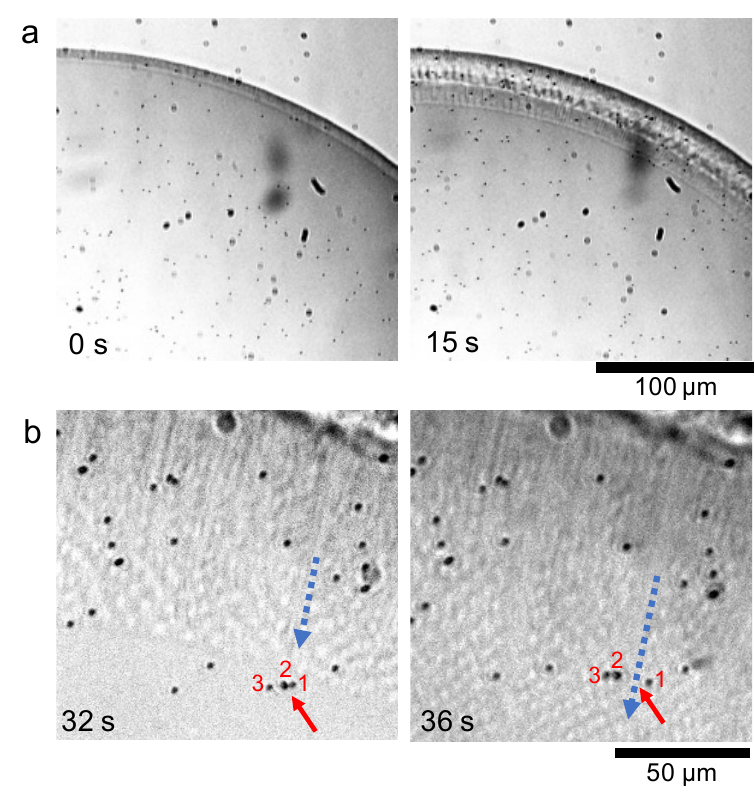}
    \caption{Dynamics of the 1 $\mu m$ sized particles inside the Pluronic droplet forming lattice pattern. a) Snapshots of the polarized images of the evaporating droplet. b) Highlighting of several particles that remain embedded in the lattice pattern during the formation of the pattern and the appearance of the radial rips. The blue dashed line represents the rips.}
    \label{fig:Fig5}
\end{figure}
 
\newpage

\newpage
\newpage
\subsection*{Supplementary Videos}
Video S1: Liquid-liquid phase separation in the Pluronic droplet 0.5\% w/w.
\\
Video S2: Lattice pattern formation in the Pluronic droplet 1\% w/w.
\\
Video S3: Closer view of the lattice pattern formation at the interface of the droplet. 
\end{document}